\documentclass[aps,pre,twocolumn,showpacs,floatfix,superscriptaddress,nofootinbib]{revtex4-2}

\usepackage{amssymb,amsmath,amstext}                
\usepackage{graphicx}                                               
\usepackage{grffile}
\usepackage{epstopdf}                                               
\usepackage{color}                                                     
\usepackage{bm}                                                        
\usepackage{appendix}                                              

\usepackage{ulem}
\usepackage{latexsym}
\usepackage[colorlinks=true,citecolor=blue,linkcolor=magenta]{hyperref}
\usepackage{cancel}
\usepackage{xcolor}
\usepackage{cleveref}
\usepackage{physics}
\usepackage{enumitem}
\usepackage{hyperref}
\usepackage{cleveref}

\newcommand{\Npsi}{\mathcal{N}_\psi}
\newcommand{\Nphi}{\mathcal{N}_\phi}
\newcommand{\psits}{\psi_{\theta}(s)}

\newcommand{\phigs}{\phi_{\gamma}(s)}
\newcommand{\phigsdag}{\phi^*_{\gamma}(s)}
\newcommand{\F}{\mathcal{F}}
\newcommand{\phipsi}{\braket{\phi}{\psi}}
\newcommand{\psiphi}{\braket{\psi}{\phi}}
\newcommand{\E}[1]{\mathbb{E}\left(#1\right)}
\newcommand{\Var}{\text{Var}~}
\newcommand{\epsmed}{\varepsilon_{\text{med}}}

\begin{document} 

\title{Fidelity and Overlap of Neural Quantum States: Error Bounds on the Monte Carlo Estimator}
\author{Tomasz Szo\l{}dra}
\affiliation{{Doctoral School of Exact and Natural Sciences, Jagiellonian University, \L{}ojasiewicza 11, PL-30-348 Krak\'ow, Poland}}
\affiliation{Instytut Fizyki Teoretycznej, 
Uniwersytet Jagiello\'nski,  \L{}ojasiewicza 11, PL-30-348 Krak\'ow, Poland}

\date{\today}

\begin{abstract}
 
 Overlap between two neural quantum states can be computed through Monte Carlo sampling by evaluating the unnormalized probability amplitudes on a subset of basis configurations. Due to the presence of probability amplitude ratios in the estimator, which are possibly unbounded, convergence of this quantity is not immediately obvious. Our work provides a derivation of analytical error bounds on the overlap in the Monte Carlo calculations as a function of their fidelity and the number of samples. Special case of normalized autoregressive neural quantum states is analyzed separately. 
\end{abstract}
\date{\today}

\maketitle

\section{Introduction}
\label{sec:intro}

Numerical simulation of many-body quantum systems on classical hardware poses a challenge due to high dimensionality of the quantum Hilbert space, growing exponentially with the system size. Exact diagonalization of many-body problems in the full Hilbert space can only be done for relatively small systems of several degrees of freedom due to memory and computational time limitations \cite{Sandvik10}. Multiple approaches to solve these problems in an approximate way have been developed to date. Among them are Quantum Monte Carlo (\textbf{QMC}) methods \cite{Hirsch82,Ceperley86, Troyer05} and tensor networks (\textbf{TN}) \cite{White92} (for reviews see \cite{Schollwock11,Orus14}), which have proven successful for certain classes of low-dimensional systems with short-range entanglement. Nevertheless, TN have limitations regarding the representable space of states and a high computational cost for tensor contractions. 

A possible way of solving the many-body problem is the use of Neural Quantum States (\textbf{NQS}), a variational neural network Ansatz for a quantum wavefunction, first introduced in \cite{Carleo17}. The neural network maps configurations into complex probability amplitudes. Due to nonlinearities in the activation functions in the network this Ansatz can represent a larger class of states than tensor networks \cite{Sharir22,Wu23} with fewer parameters (see, however, \cite{Lin22,Passetti23} and in particular \cite{Denis23}). Parameters of the network are optimized variationally by Monte Carlo (\textbf{MC}) sampling of the configurations (see e.g. \cite{Dawid22} for a pedagogical introduction). The difficulty now shifts towards costly MC sampling and navigation on the complicated optimization landscape of parameters, rather than mere expressibility of the Ansatz. Nevertheless, it has been demonstrated that NQS can be used to find ground states of 1- and 2-dimensional spin systems \cite{Sharir20, Hibat20}, perform unitary time evolution \cite{Carleo17, Carleo17u, Schmitt20}, learn physical wavefunction from experimental data \cite{Torlai18} or simulate open systems in contact with reservoir \cite{Yoshioka19, Carrasquilla19, Vicentini19, Hartmann19, Luo21, Reh21}.

Quantum fidelity and overlap are fundamental quantities used for measuring the distance between quantum states. They can be used e.g. to iteratively find excited states starting from the groundstate (see \cite{Choo18} for excited state calculation in NQS), map out phase diagrams \cite{Zhou08, Damski15}, or to verify how far the quantum state prepared in the lab is from the desired one \cite{Flammia11}. In the context of approximate numerical Ans\"{a}tze, such as NQS, calculation of fidelity between the approximate and exact state, although accessible only for small systems, gives a quantitative estimate on the quality of the Ansatz. It is also a useful tool in diagnosis of the unitary time evolution algorithms, as performed in \cite{Hofmann22}. Not all numerical methods of simulating many-body systems allow for a direct, efficient calculation of fidelity - see eg. algorithm introduced in \cite{Zhou08} for 2-dimensional tensor networks.

Although the calculation of fidelities has already been successfully performed with NQS \cite{Choo18,Hofmann22,Sinibaldi23} in the standard Monte Carlo sampling fashion, and it has recently been rigorously proven it \textit{can} be computed with a finite number of samples \cite{Havlicek23}, in this paper we perform a detailed study of the error of such an estimate. Our calculations yield simple, analytical error bounds on both fidelity and overlap calculated with NQS via Monte Carlo sampling, which in some settings are tighter than those of Ref.~\cite{Havlicek23}.

This paper is organized as follows. In Section \ref{sec:main} we establish the notation and pedagogically revise how one calculates the overlap and fidelity for neural quantum states through Monte Carlo sampling. Section \ref{sec:error_bounds} is devoted to the derivation of the error bounds for both of these quantities. We separately treat the case of normalized states, eg. autoregressive neural networks (\textbf{ARNN}), for which sampling from only a single network is required, and the general case of unnormalized probability amplitudes. In Section \ref{sec:benchmarks} we numerically benchmark the error bounds for a physical system. We conclude in Section \ref{sec:summary}.


\section{Overlap and fidelity}
\label{sec:main}
\begin{figure}
    \centering
    \includegraphics[width=\columnwidth]{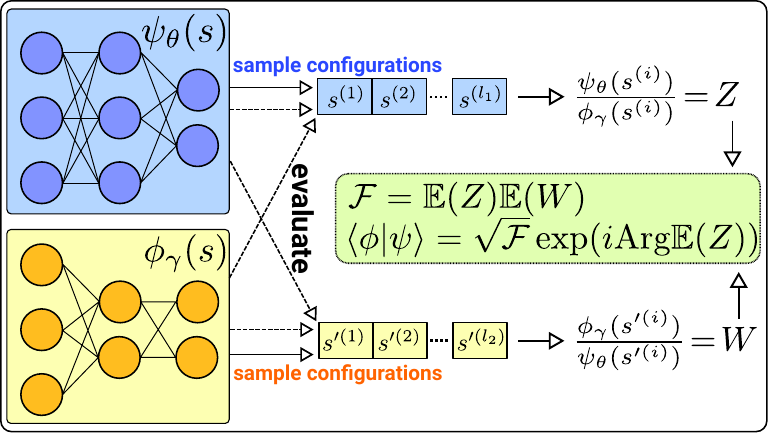}
    \caption{Scheme of the fidelity and overlap measurement algorithm. If both states are normalized, e.g., they are represented by an autoregressive Ansatz, sampling from only one of the NQS becomes sufficient, see text.}
    \label{fig:scheme}
\end{figure}
Assume that we are given two NQSs, parametrized by variational parameters $\theta, \gamma$, respectively, that encode approximations of two quantum states
\begin{eqnarray}
    \ket{\psi} &=& \sum_{s} \frac{\psits}{\sqrt{\Npsi}} \ket{s}, \quad \Npsi \equiv \braket{\psi},\\
    \ket{\phi} &=& \sum_{s} \frac{\phigs}{\sqrt{\Nphi}} \ket{s}, \quad \Nphi \equiv \braket{\phi},
\end{eqnarray}
where $\lbrace \ket{s} \rbrace_s$ are basis vectors of the Hilbert space with $\braket{s}{s'} = \delta_{ss'}$. For example, for $L$ spins-1/2, $s=s_1\dots s_L$ with $s_i=\uparrow, \downarrow$ and the number of basis elements is $2^L$. We consider a general case of NQS with $\Npsi, \Nphi$ not necessarily equal to $1$. Some of our results obtained for $\Npsi=\Nphi=1$, e.g. for ARNN, are analyzed separately.
We calculate the overlap
\begin{equation}
    \phipsi = \sum_s \frac{\psits \phigsdag}{\sqrt{\Npsi \Nphi}}
    \label{eq:phipsi}
\end{equation}
and fidelity $\F = \left| \phipsi \right|^2$. As we are dealing with NQS, we can (i) evaluate amplitudes $\psits$, $\phigs$ efficiently given configuration $s$, (ii) generate $n$ samples of configurations $\lbrace s^{(1)} \dots s^{(n)} \rbrace$ coming from probability distributions $\left| \psits \right|^2 / \Npsi$ and $\left| \phigs \right|^2 / \Nphi$.

Although terms \eqref{eq:phipsi} are accessible, naive evaluation of the network on all basis elements is intractable for large systems due to the dimension of the Hilbert space growing exponentially with the system size. The normalization factors $\Npsi, \Nphi$ also require a summation of exponentially many terms. However, one can use a standard trick for the MC calculation of operator expectation values with NQS \cite{Carleo17}, i.e., rewrite Eq.~\eqref{eq:phipsi} as:
\begin{equation}
    \phipsi = \sqrt{\frac{\Nphi}{\Npsi}} \sum_s \frac{\left|\phigs\right|^2}{\Nphi} Z(s)
    \label{eq:trick}
\end{equation}
where $Z(s) = \psits /\phigs$. Then $\phipsi = \sqrt{\frac{\Nphi}{\Npsi}} \E{Z(s)}$, with configurations $s$ drawn from the probability distribution $\left| \phigs \right|^2 / \Nphi$. 
To estimate $\E{Z}$ one generates $n_1$ samples $\lbrace s^{(1)}, \dots s^{(n_1)} \rbrace$ according to this probability distribution and constructs an empirical MC estimator

\begin{equation}
    Y_1 = \frac{1}{n_1} \sum_{i=1}^{n_1} Z(s^{(i)}).
    \label{eq:Y1}
\end{equation}
It is clear that $\E{Y_1} = \E{Z}$.

If both states are normalized with $\Npsi=\Nphi=1$, e.g. they are represented by an autoregressive Ansatz, one can immediately read off the overlap $\phipsi = \E{Y_1}$. 

However, for unnormalized states, the preceding factor of $\sqrt{\Nphi/\Npsi}$ remains unknown.
This issue can be circumvented by computing
\begin{eqnarray}
\psiphi  =  \sqrt{\frac{\Npsi}{\Nphi}} \sum_s \frac{\left| \psits \right|^2}{\Npsi} W(s),
\end{eqnarray}
where $W(s) = \phigs / \psits$ and we used the same trick as in Eq.~\eqref{eq:trick} but for $\psits$.
In the similar way as before, we can generate $n_2$ samples $\lbrace s'^{(1)}, \dots s'^{(n_2)} \rbrace$ but now from
from $\left| \psits \right|^2/\Npsi$ and define an estimator 
\begin{equation}
     Y_2 = \frac{1}{n_2} \sum_{i=1}^{n_2} W(s'^{(i)}).
\end{equation}
Of course, $\E{Y_2} = \E{W}$, and fidelity reads
\begin{eqnarray}
    \F & = & \phipsi \psiphi = \sqrt{\frac{\Nphi}{\Npsi}} \sqrt{\frac{\Npsi}{\Nphi}}\E{Y_1} \E{Y_2}\\\nonumber
    & = & \E{Y_1 Y_2},
\end{eqnarray}
i.e., normalization factors cancel out, while $\E{Y_1} \E{Y_2} = \E{Y_1 Y_2}$ as variables $Y_1$, $Y_2$ are independent because they come from two different random processes. Using this result for the overlap, we can write
\begin{equation}
    \phipsi = \sqrt{\left|\E{Y_1 Y_2}\right|} e^{i \alpha}, \quad \alpha = \arg \E{Y_1}.
    \label{eq:phipsi_MC}
\end{equation}
Note that the definition of the complex phase $\alpha$ involves only $Y_1$ because a "symmetric" formula ${\alpha=\left(\arg \E{Y_1} -  \arg \E{Y_2}\right) / 2}$ would be defined up to mod $\pi$.

\section{Error bounds}
\label{sec:error_bounds}
We provide an upper bound on the number of samples $n_1, n_2$ (or only $n_1$ for the case of normalized states) required to estimate $\phipsi$ and $\F$ with a fixed additive error and failure probability $\delta$, see Fig.~\ref{fig:error_bounds}. The bounds are calculated using the framework first employed in \cite{Flammia11} in the context of measuring fidelities in quantum circuits through Pauli measurements.

\subsection{Normalized states}
\label{sec:normalized}
For simplicity, we first consider the case of normalized states $\Npsi=\Nphi=1$, e.g. encoded using ARNN architectures, which generate configurations site-by-site by implementing conditional probabilities $P(s_i|s_{i-1}, \dots s_1)$ \cite{Wu19},

\begin{equation}
    P(s_1, s_2, \dots, s_L) = \Pi_{i=1}^L P(s_i|s_{i-1}, \dots s_1).
\end{equation}

\begin{figure}
    \centering
    \includegraphics[width=.45\columnwidth]{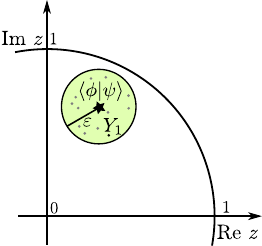}
    \caption{Error bounds of $\phipsi$ in case $\Npsi=\Nphi=1$, eg. for autoregressive NQS. a) Star represents the true value of $\phipsi$ which can be calculated if all amplitudes of both wave functions $\psits$, $\phigs$ are known. Estimates $Y_1$, represented by dots, fall into the circle of radius $\varepsilon$ around $\phipsi$ with probability $1-\delta$. 
    }
    \label{fig:error_bounds}
\end{figure}

We use the Chebyshev's inequality 

\begin{equation}
    \Pr \left( \left| Y_1 - \E{Y_1} \right| \geq \sigma k \right) \leq \frac{1}{k^2}, \quad \sigma = \sqrt{\Var Y_1},
    \label{eq:chebyshev}
\end{equation}
which holds for any probability distribution of the random variable $Y_1$ with finite first and second moments. It states that the probability of a single realization of $Y_1$ achieving a value of $k$ standard deviations away from the mean is smaller than $1/k^{2}$. In our case, ${\Var Y_1 = (\Var Z) / n_1}$, see Eq.~\eqref{eq:Y1}, while
\begin{eqnarray}
\label{eq:varZ}
\Var Z &=& \E{ZZ^*} - \E{Z}\E{Z^*} \\\nonumber
&=& \sum_s \left|\phigs\right|^2 \frac{\left|\psits\right|^2}{\left|\phigs\right|^2} - \F = 1-\F,
\end{eqnarray}
where we used $\phipsi=\E{Z}$ and normalization of $\ket{\psi}$. Setting $k=1/\sqrt{\delta}$ we obtain
\begin{equation}
    \Pr \left( \left| Y_1 - \phipsi \right| \geq \varepsilon\right) \leq \delta,
    \label{eq:error_arnn}
\end{equation}
where 
\begin{equation}
    \varepsilon = \sqrt{\frac{1-\F}{n_1 \delta}}.
\end{equation}
That is, without any prior knowledge about the fidelity $\F$, assuming the worst-case scenario of orthogonal states with $\F=0$, we need at least $n_1 = 1/ (\varepsilon^2 \delta)$ samples from $\left|\phigs\right|^2$ in order for the estimate $Y_1$ to fall within a complex disk of radius $\varepsilon$ around the true value $\phipsi$ with probability $1-\delta$, see Fig.~\ref{fig:error_bounds}a. Notice that the required number of samples is independent of any properties of the state, system size and any system details such as local Hilbert space and spatial dimension. 
We also note that although the Central Limit Theorem guarantees that the distribution of $Y_1$ asymptotically converges to the normal distribution on the complex plane ${N\left(\mu=\braket{\phi}{\psi},~\sigma=\sqrt{(1-\F)/n_1} \right)}$ for $n_1\rightarrow \infty$, the rate of convergence with $n_1$ depends on the details of the probability distribution of $Z(s^{(i)})$ and should be computed separately for each pair of quantum states $\ket{\psi}$, $\ket{\phi}$. On the contrary, the Chebyshev's inequality holds without the assumption of the distribution of $Y_1$.

The fidelity could in principle be estimated by ${\F = \E{Y_1 Y_1^*}}$. We were unable to provide bounds for its variance as this expression contains terms $\E{(ZZ^*)^2}$ and $\E{ZZ^*Z}\E{Z^*}$ which do not simplify as in Eq.~\eqref{eq:varZ}. On the other hand, the interval in which we expect to find the estimate of $\F$ follows directly from the error in overlap,

\begin{equation}
    \Pr \left( \left| |Y_1|^2 - \F \right| \geq 2\varepsilon\sqrt{\F} + \varepsilon^2\right) \leq \delta,
    \label{eq:error_arnn_F}
\end{equation}
where $\varepsilon$ was previously defined in Eq.~\eqref{eq:error_arnn}. This is obtained from 
\begin{eqnarray}
    ||Y_1|^2-\F| &=& ||(\phipsi + e^{i\varphi} \varepsilon)|^2-\F|\\\nonumber
    &=&\varepsilon| 2 \Re\left(\phipsi e^{-i\varphi}\right) + \varepsilon| \\\nonumber
    &\leq& \varepsilon (2 \sqrt{\F} + \varepsilon)
\end{eqnarray}
where we used $-\sqrt{\F} \leq \Re \left(\phipsi e^{-i\varphi}\right) \leq \sqrt{\F}$ and $e^{-i\varphi}$ is an arbitrary complex phase.

For large $n_1 \delta \gg 1$ the fidelity error attains a maximum at $\F=0.5$. Minima of the error occur at $\F=0$ and $\F=1$. At $\F=1$ the error reaches $0$, while for $\F=0$ it converges to $1/(n_1 \delta)$.

\subsection{General case}
\label{sec:general}
The same technique is directly generalized to unnormalized states. 
First, we derive an error bound for the MC estimate $Y_1 Y_2$ of fidelity $\F$ through Chebyshev's inequality \eqref{eq:chebyshev}. Using the independence of the variables $Y_1$, $Y_2$, and Eq.~\eqref{eq:varZ}, we find that
\begin{equation}
    \Var Y_1 Y_2 = \F (1 - \F) \frac{n_1+n_2}{n_1 n_2} + \frac{(1-\F)^2}{n_1 n_2}.
    \label{eq:variance_mcmc}
\end{equation}
Full derivation is presented in the Appendix \ref{app:A}. 

Given the total number of samples $n=n_1+n_2$, choosing equal sampling $n_1=n_2=n/2$ minimizes the variance. Plugging $k=1/\sqrt{\delta}$ into the Chebyshev's inequality for $Y_1 Y_2$ in Eq. \eqref{eq:chebyshev}, and using the variance formula \eqref{eq:variance_mcmc}, for the optimal choice of $n_1=n_2=n/2$ samples we obtain
\begin{equation}
    \Pr \left( \left| Y_1 Y_2 - \F \right| \geq \varepsilon ' \right) \leq \delta,
    \label{eq:error_mcmc}
\end{equation}
where
\begin{equation}
    \varepsilon ' = 2\sqrt{\frac{ \F(1-\F) + \frac{(1-\F)^2}{n}}{n \delta}}.
    \label{eq:epsilonprime}
\end{equation}
For a large number of samples $n\gg 1$, the minimal error $\varepsilon '$ is obtained for $\F=0$ and $\F=1$, while ${\F=1/2}$ maximizes the error. Let us compare this result to the case of single-state sampling with the same number $n_1=n$ of samples. 

Taylor expansion of $\varepsilon'$ as a function of $1/n$ around zero gives 
\begin{equation}
    \varepsilon' \approx 2\sqrt{\frac{\F(1-\F)}{n\delta}} + \frac{(1-\F)^2}{n^2\delta}.
    \label{eq:epsilonprime2}
\end{equation}
This means that the leading term is the same as in Eq.~\eqref{eq:error_arnn_F}, while a higher-order correction to the error is smaller here by a factor of $\sim n/(1-\F)$. However, in practice one always uses $n\gg 1$ so there are no advantages of using one or the other method, irrespective of the fidelity value $\F$.

Our error bound can be compared to that of Ref.~\cite{Havlicek23}, where an robust estimator based on the calculation of the median of multiple estimates of means of $Y_1 Y_2$ is considered. With a total of $n$ samples, i.e. $n$ queries of the amplitude ratios, the fidelity is estimated up to $\epsmed$ additive error with failure probability $\delta=\exp\left(-n \epsmed^2 / 64\right)$. It is straightforward to show that at the same failure probability $\delta$, our error bound given by Eq.~\eqref{eq:epsilonprime} is smaller, $\varepsilon' < \epsmed$, if the condition
\begin{equation}
    \F (1-\F) + \frac{(1-\F)^2}{n} < -16~\delta \ln{\delta}
\end{equation}
is met. In the limit of large $n\gg 1$, this is always true if $0.00263 \lessapprox \delta \lessapprox 0.984)$ since the l.h.s. is always less or equal $0.25$. The inequality is also satisfied in the interesting cases when states are either close to each other, $\F\approx 1$, for any $n$, or when they are orthogonal, $\F=0$, for $n$ dependent on the chosen $\delta$.

Having the fidelity interval, we can read off the error bounds for the overlap defined by Eq.~\eqref{eq:phipsi_MC}. Since, with probability $1-\delta$, $|Y_1 Y_2|$ lies in the interval $[\F-\varepsilon ', \F + \varepsilon ']$, the estimator $\sqrt{|Y_1 Y_2|}$ for $\left|\phipsi\right|$ lies in the interval $[\min (0, \sqrt{\F-\varepsilon '}), \max(1, \sqrt{\F + \varepsilon '}]$ with the same probability.

\begin{figure}
    \centering
    \includegraphics[width=.5\columnwidth]{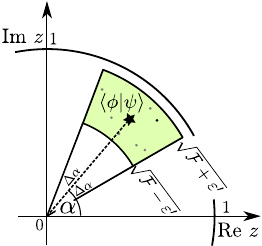}
    \caption{Error bounds in the general case of unnormalized states and two sets of samples. Estimate Re$Y_1 Y_2$ of true fidelity $\F$ falls into the interval $[\F-\varepsilon ', \F+\varepsilon ']$ with probability at least $1-\delta$. Estimate \eqref{eq:phipsi} of overlap $\phipsi$ falls into the shaded area above with the same probability.}
    \label{fig:error_bounds2}
\end{figure}

The error in the complex phase $\alpha$ is due to the uncertainty of the phase in $Y_1$. With probability $1-\delta$, $Y_1$ lies in a disk of radius $r= \varepsilon' \sqrt{\Npsi/\Nphi}$ with a center at $\sqrt{\Npsi/\Nphi}\phipsi$. The disk spreads over complex angles $[\alpha-\Delta\alpha, \alpha + \Delta\alpha]$, and simple trigonometry gives
\begin{eqnarray}
    \Delta \alpha &=& \arcsin \frac{r}{\sqrt{\Npsi/\Nphi}\left|\phipsi\right|} \\\nonumber
    &=&\arcsin \frac{2\sqrt{\F(1-\F)+\frac{(1-\F)^2}{n}}}{\sqrt{\F n\delta }}.
    \label{eq:error_mcmc_phase}
\end{eqnarray}
(If the argument under $\arcsin$ is greater than $1$, we separately define the angle $\Delta\alpha$ as $\pi$, so that $\alpha$ remains undefined). Thus, with probability at least $1-\delta$, the MC estimator of $\phipsi$ lies in the region presented in Fig.~\ref{fig:error_bounds2}.

\section{Numerical benchmarks}
\label{sec:benchmarks}
\subsection{Method}
We numerically validate the analytical error bounds in the following way. To simulate a generic, entangled state $\ket{\psi(0)}$ we consider a 1-dimensional chain of $L$ spins-1/2 and a neural network variational Ansatz with random weights (see eg. \cite{Xiaoqi22} for the properties of restricted Boltzmann machine NQS with Gaussian weights). In order to measure the error as a function of fidelity, we scan a range of fidelities by evolving this state in time to $\ket{\psi(t)} = \exp(-i H t) \ket{\psi(0)}$ with the Hamiltonian of the transverse field Ising model,
\begin{equation}
    H = -\sum_{i=1}^L J \sigma_i^z \sigma_{i+1}^z + g \sigma_i^x,
\end{equation}
with periodic boundary conditions ${\sigma_{L+1}^z \equiv \sigma_{1}^z}$, ${J=1}$ and ${g=-0.5}$. Time evolution up to time $t=2$ is performed using the time-dependent Variational Monte Carlo (tVMC) method \cite{Carleo17, Yuan19,Schmitt20,Gutierrez22}. Our codes are available at \url{github.com/tszoldra/jVMC_overlap}. We use the TDVP \cite{Haegeman11} algorithm, implemented in jVMC 1.2.4 \cite{Schmitt22,Schmitt22code} framework, with Euler integrator with timestep $0.001$, regularization cutoff in the pseudoinverse of the quantum Fisher matrix $\epsilon_{\text{pinv}}=10^{-6}$ and $8192$ MC samples for each timestep (see \cite{Schmitt22} for details on regularization techniques). We do not care how well the variational Ansatz approximates the actual time-evolved  physical wave function as we are only interested in the decrease of fidelity between the time-evolved and the initial state. 

The error of the overlap estimation is measured by comparing the Monte Carlo estimate with the exact summation over all $2^L$ basis states as in Eq.~\eqref{eq:phipsi} which could be done for the moderate system sizes $L \leq 32$. Cases of autoregressive and unnormalized neural networks are analyzed separately below. All results are produced using $n=65536$ samples (in ARNN case, $n_1=n$, in general case $n_1=n_2=n/2$) and are averaged over $10$ random initial states (random weights of the neural network) and $100$ independent MC calculations of the overlap or fidelity. Error bars denote an estimate of one standard deviation for a single sample, i.e., a deviation one may expect when computing a single estimate of the overlap or fidelity using $n$ samples in total.

\begin{figure}
    \centering
    \includegraphics[width=0.7\columnwidth]{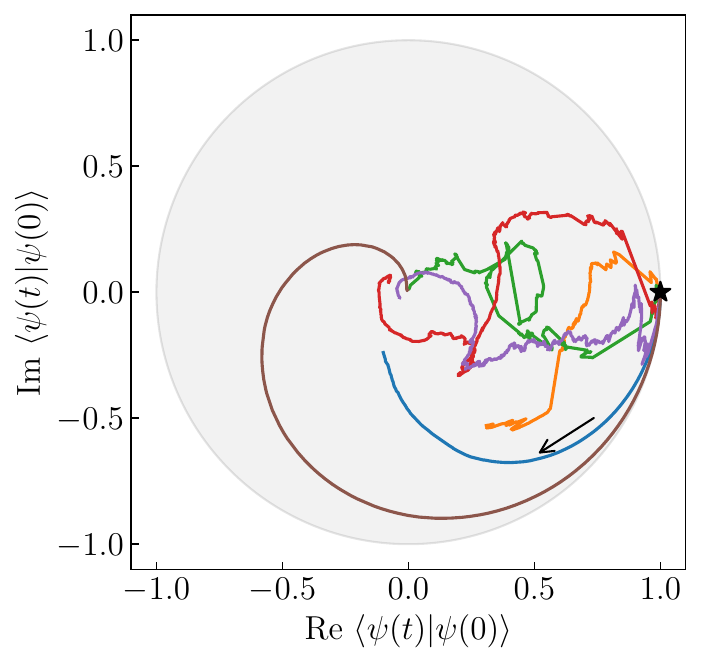}
    \caption{Examples of time evolution of the exact overlap $\braket{\psi(t)}{\psi(0)}$ from $6$ different initial states (different colors) and the autoregressive recurrent neural network. Similar plots would be obtained for general unnormalized NQS. This gives us access to states with different fidelities controlled by time $t$. Arrow denotes increasing time for one trajectory. All trajectories start at time $t=0$ denoted by a star.}
    \label{fig:overlap_evol}
\end{figure}

Our ARNN Ansatz is a densely connected recurrent neural network \cite{Hibat20} with hidden size $16$ and depth $4$, yielding $1090$ real variational parameters in total, and an exponential linear unit (ELU) \cite{Clevert16} activation function. Random initial weights come from a uniform distribution on the interval $[0,1]$, rescaled by the square root of the average number of inputs and outputs in a given layer. Due to the autoregressive property, MC sampling of spin configurations from this network is exact, i.e. MC samples are perfectly uncorrelated. Fig.~\ref{fig:overlap_evol} shows exemplary time evolutions of an exact overlap $\braket{\psi(t)}{\psi(0)}$ on the complex plane.

The unnormalized Ansatz is a restricted Boltzmann machine (\textbf{RBM}) \cite{Carleo17,Melko19} with complex parameters, hidden size of $32$ units, and no bias term. Number of complex variational weights varies between system sizes and is equal to $32L$. Random initial weights have real and complex parts uniformly distributed on the interval $[0,0.01]$. We sample configurations from this NQS using Markov chain Monte Carlo Metropolis-Hastings algorithm with random spin flip as a step proposal, $L$ Monte Carlo steps for a single sweep, and $25$ sweeps for an initial "burn-in"/thermalization.

\subsection{Results}
\begin{figure}
    \centering
    \includegraphics[width=.85\columnwidth]{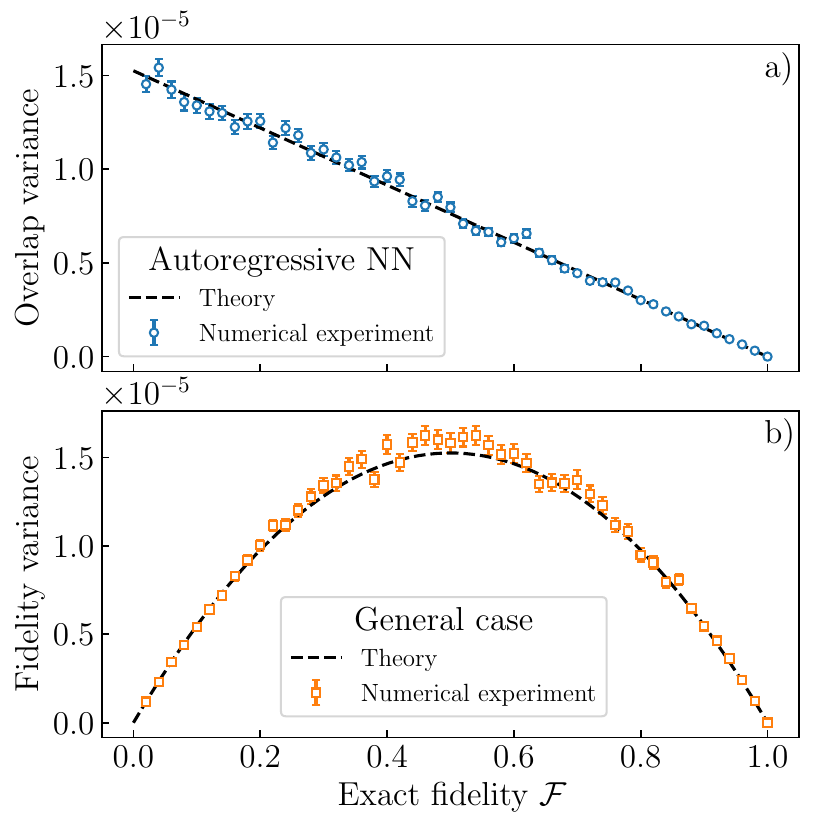}
    \caption{Numerical estimates of variances of a) overlap in the autoregressive NN calculation, b) fidelity in the general case. Computed variances closely follow the theoretical predictions from Eqs.~\eqref{eq:varZ} and \eqref{eq:variance_mcmc}, respectively, in both panels. We used system size $L=24$ and $n=65536$ MC samples.}
    \label{fig:variances}
\end{figure}

In FIG.~\ref{fig:variances} we verify the the variance formulae Eq.~\eqref{eq:varZ} for ARNN (panel a) and Eq.~\eqref{eq:variance_mcmc} for unnormalized NQS (panel b) for system size $L=24$. In both cases, we see a very good agreement between the theoretical prediction and the numerical result within error bars. Here, unlike in all other plots, the error bars are the standard deviations from the mean value, i.e. error of the mean estimate, not a deviation expected for a single MC estimate.
\begin{figure*}
    \centering
    \includegraphics[width=\textwidth]{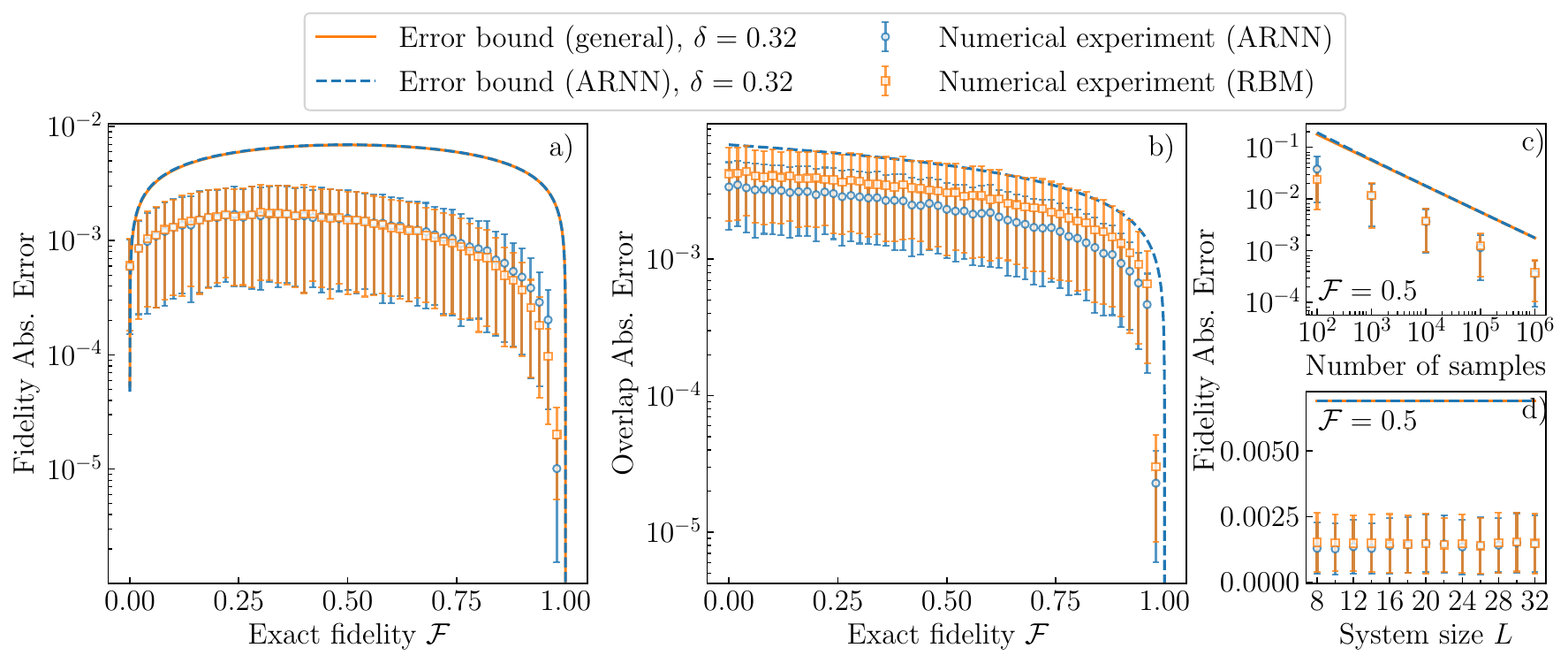}
    \caption{a) Absolute error of the fidelity estimator with respect to the exact value given by full basis summation as a function of the fidelity $\F$ for $n=65536$ MC samples. The analytical error bounds at a failure rate of $\delta=0.32$ are denoted by solid line (unnormalized NQS) and dashed line (autoregressive NQS). b) Overlap absolute error for the same system. c) Error as a function of the number of samples for the worst-case scenario $\F=0.5$. The error decreases as $\sim 1/\sqrt{n}$. d) Independence of the error and the error bound on the system size $L=8\dots32$. Error bounds are vaild for arbitrary finite $L$. All results are averaged over $10$ initial states and $100$ Monte Carlo calculations.}
    \label{fig:error_abc}
\end{figure*}

We then verify the error bounds. FIG.~\ref{fig:error_abc}a) shows mean absolute errors of fidelity MC estimate for ARNN and RBM networks for system size $L=24$. Since the error depends on $\F$, interval of fidelities $\F\in [0, 1]$ has been divided into $50$ bins of size $0.02$ each. The errors are contrasted with theoretical bounds \eqref{eq:error_arnn_F} and \eqref{eq:error_mcmc} at failure probability $\delta=0.32$, corresponding to deviation larger than $1\sigma$ for a Gaussian variable. It is clear that both error bounds resulting from the Chebyshev's inequality are overestimated. Similar conclusions apply to complex-valued overlap absolute errors in FIG.~\ref{fig:error_abc}b) for which error bounds are also not saturated.

In FIG.~\ref{fig:error_abc}c) we investigate the mean fidelity error as a function of the total number of samples $n$ for the worst-case fidelity $\F=0.5$ and the largest system size ${L=32}$. In practice, the plot is produced for fidelities from a certain range around this value, $\F\in[0.45,0.55]$. For both ARNN and RBM the error decreases as $1/\sqrt{n}$.

Finally, in FIG.~\ref{fig:error_abc}d) we demonstrate that the fidelity estimation error is independent of the system size $L$. Error does not change to within the error bar between $L=8$ and $L=32$. The analytical error bound guarantees that fidelity is also computed with a prescribed accuracy and failure probability for larger systems.

\section{Summary}
\label{sec:summary}

In this Article we studied the convergence properties of the Monte Carlo estimate of the quantum overlap and fidelity. By analytically computing the variance of the MC estimator and utilizing the Chebyshev's inequality, we were able to show that the error is explicitly given only by a total number of Monte Carlo samples and fidelity between the quantum states in consideration. We have verified our bounds on the error by comparing the resulting fidelity with the full summation over the Hilbert space up to dimension $2^{32}\approx 4.3 \cdot 10^9$. 

In the case of autoregressive states, only one of the NQS can be sampled from for the overlap estimation. This feature can be used for example if one wants to compute fidelity between an autoregressive NQS and a Matrix Product State and sample only the former to possibly avoid expensive tensor network contractions (see also \cite{Sandvik07} for Monte Carlo sampling of tensor networks). 

We identified two regimes in which the error of fidelity estimate goes down. A small error for orthogonal states around $\F\approx 0$ may explain the remarkable efficiency of the excited states search algorithm \cite{Choo18}. A small error at $\F\approx1$ is also a good sign for a class of algorithms that are supposed to variationally optimize states with low infidelity. 

\textit{Note} In the late stage of preparation of this manuscript, I became aware of recently published Ref.~\cite{Sinibaldi23}. Authors estimate the infidelity of NQS using a technique of Control Variates which gives a lower relative MC error than a standard estimate when operating around $\F\approx 1$. Better convergence of this estimate is used to construct an efficient \textit{projected} tVMC unitary time evolution algorithm. Authors demonstrate that the natural measure of the MC estimate quality is the signal-to-noise ratio, defined as the quantity of interest in units of its standard deviation. It is shown that a direct MC estimate of infidelity $1-\F$ close to $0$ through an algorithm described here would require a diverging number of $n\sim 1/(1-\F)$ samples in order to conserve the signal-to-noise ratio and have a sufficient accuracy for performing the variational minimization. Authors also derive a variance formula which is a special case of  Eq.~\eqref{eq:variance_mcmc} in the limit $\F \rightarrow 1$, $n \gg 1$.

\acknowledgements
I thank J. Zakrzewski, P. Sierant, M. Lewenstein and P. Korcyl for inspiring discussions. This work has been realized within the Opus grant 2019/35/B/ST2/00034, financed by the National Science Centre (Poland). I gratefully acknowledge Poland’s high-performance computing infrastructure PLGrid (HPC Centers: ACK Cyfronet AGH) for providing computer facilities and support within computational grant no. PLG/2022/015986.

%

\appendix

\section{Calculation of the fidelity variance}
\label{app:A}
Variance of fidelity estimate $Y_1 Y_2$ is defined as 
\begin{equation}
    \Var Y_1 Y_2 = \E{Y_1 Y_2 Y_1^* Y_2^*} - \E{Y_1 Y_2} \E{Y_1^* Y_2^*}.
\end{equation}
Using the independence of random variables $Y_1, Y_2$, this boils down to 
\begin{eqnarray}
    \Var Y_1 Y_2 &= &\E{Y_1 Y_1^*} \E{Y_2 Y_2^*} \\
    &&- \E{Y_1}\E{Y_1^*}\E{Y_2}\E{Y_2^*}.
\end{eqnarray}
Notice that
\begin{equation}
    \Var Y_1 = \E{Y_1 Y_1^*} - \E{Y_1}\E{Y_1^*} = \frac{\Var Z}{n_1}
\end{equation}
and for unnormalized states
\begin{equation}
    \Var Z  = \frac{\Npsi}{\Nphi}(1-\F),
\end{equation}
see Eq.~\eqref{eq:varZ} in the main text. Since $\E{Y_1}=\E{Z}=\frac{\Npsi}{\Nphi}\phipsi$, we get
\begin{eqnarray}
    \E{Y_1 Y_1^*}&=&\E{Y_1}\E{Y_1^*} + \frac{\Var Z}{n_1}  \\\nonumber
     & = & \frac{\Npsi}{\Nphi}\left(\F + \frac{1-\F}{n_1}\right).
\end{eqnarray}
Similarly,
\begin{equation}
    \E{Y_2 Y_2^*} = \frac{\Nphi}{\Npsi}\left(\F + \frac{1-\F}{n_2}\right).
\end{equation}
Thus, 
\begin{eqnarray}
    \Var Y_1 Y_2 &=& \left(\F + \frac{1-\F}{n_1}\right)\left(\F + \frac{1-\F}{n_2}\right)\!-\!\F^2 \\\nonumber
     & = & \F(1-\F) \frac{n_1 + n_2}{n_1 n_2} + (1-\F)^2 \frac{1}{n_1 n_2}.
\end{eqnarray}
Both terms in the variance are mimimized by choosing an equal number of samples $n_1=n_2=n/2$.

\end{document}